\documentclass[twoside]{physcon17_onecolumn}

\usepackage{graphicx,array}
\usepackage{amssymb,amsmath}
\usepackage{hyperref}
\usepackage{balance}

\usepackage[T1]{fontenc}
\usepackage[utf8]{inputenc}

\title{Complex dynamics in a vehicle platoon with nonlinear drag and ACC controllers}

\author{ \twlsfb Giacomo Innocenti
    \affiliation{
      Dipartimento di Ingegneria dell'Informazione\\
      Università  degli Studi di Firenze\\
      Italy\\
      giacomo.innocenti@unifi.it
    }
    \and \twlsfb Michele Basso
    \affiliation{
      Dipartimento di Ingegneria dell'Informazione\\
      Università  degli Studi di Firenze\\
      Italy\\
      michele.basso@unifi.it
    }
}

\begin{document}
\maketitle
\begin{abstract}
In this paper a novel platoon model is presented.
Nonlinear aerodynamic effects, such as the wake generated by the preceding vehicle, are considered, and their influence in the set up of a Adaptive Cruise Controller (ACC) is investigated.
To this aim, bifurcation analysis tools are exploited in combination with an embedding technique independent from the vehicles number.
The results highlight the importance of a proper configuration for the ACC in order to guarantee the platoon convergence to the desired motion.
\end{abstract}
\section{Introduction}
\begin{figure}[t]
\begin{center}
\includegraphics[width=0.70\columnwidth]{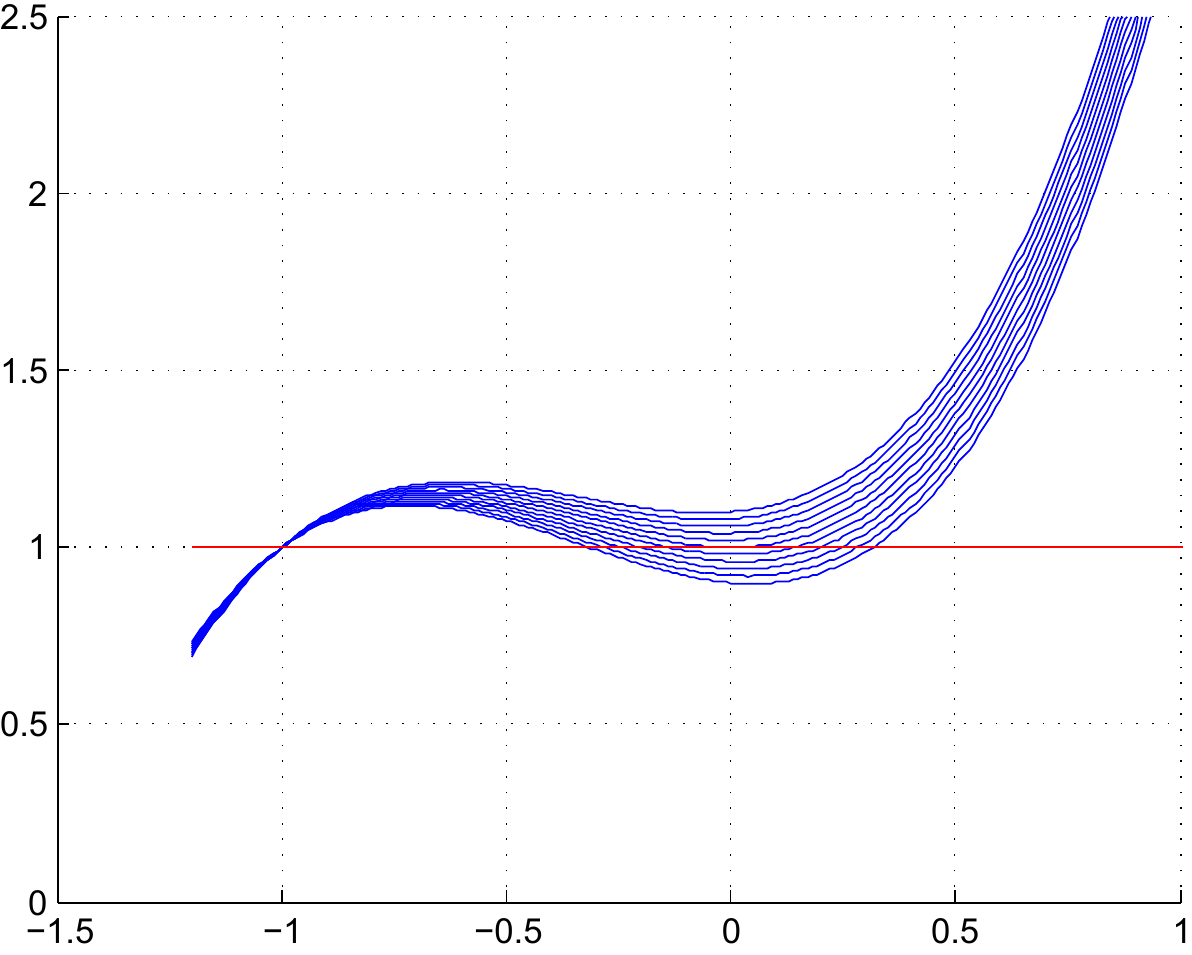}
\end{center}
\caption{The nonlinear drag for different values of the inter-vehicle distance from the preceding one for $\alpha=1$, $\beta=0.1$, and $\gamma=1$.
\label{fig:drag}} 
\end{figure}
In the recent years the technological advances in autonomous/self-driven vehicles have focus the attention also on the problem of formation control.
In this framework, terrain vehicles platoons have received great interest from the scientific community and many projects have already been funded to improve their management technologies~\cite{SARTRE}.
One of the oldest and most effective control strategies to enforce a certain formation in a vehicle platoon is the Adaptive Cruise Control~\cite{Konrad}.
However, despite its successful implementation and the numerous variants, there is no general analytic method to set it up for a large number of units.
Moreover, ACC is commonly referred to the linearized problem, and it is still not clear how possible nonlinear effects may affect the overall behaviour.

In this paper we introduce a novel model for a simple platoon featuring a nonlinear drag, that takes into account also the effects due to the wake generated by the preceding vehicle.
Such a model will be investigated in a classical nonlinear bifurcation analysis framework by transforming the problem via a recent embedding technique independent from the vehicle number.
The aim is to conceive a simple though effective approach able to provide qualitative tools for inferring whether a certain configuration of the ACC controller may cause the rise of complex phenomena, such as, for instance, traveling waves.
\section{Platoon model}
Hereafter, we will consider a 1D platoon of identical vehicles moving along a closed path with no intersections, i.e. a circuit.
The platoon is supposed to be in jam condition, i.e. the first unit sees the last one ahead of itself.
Let us consider for each vehicle a simple, though widespread in the literature, model of the form (see, e.g.,~\cite{Kwon} and the references therein):
\begin{align}
\label{eq:the model}
	a_i =&~ \frac{1}{m}\big(u_i-f(s_{i-1}-s_i,v_i)\big) ~,
\end{align}
where $s_i$, $v_i=\dot{s}_i$, and $a_i=\ddot{s}_i$ are respectively its absolute position along the circuit, the speed, and the acceleration.
In~\eqref{eq:the model} $m$ stands for the vehicle's mass, $u_i$ is its control input\footnote{Here we neglect the actuator dynamics, so the control acts at the same level of the acceleration.}, and $f$ is a nonlinear drag depending on both the speed and the distance from the preceding vehicle:
\begin{align*}
	f(s_{i-1},s_i,v_i) =&~ \alpha v_i\big((v_i-\nu)^2-\beta(\Delta s_i-\mu)\big)+\gamma ~,
\end{align*}
where $\Delta s_i=s_{i-1}-s_i$.
Drag $f$ accounts for a constant friction component, namely $\gamma>0$, and the friction depending from air resistance, that is assumed to grow as the cube of the speed according to coefficient $\alpha>0$.
This latter component of the drag is also supposed to be lightened by the presence of the preceding vehicle as a consequence of the wake, whose effects, tuned by $\beta>0$, are represented by a local minimum at $v_i=\nu$ when the inter-vehicular distance is at $\Delta s_i=\mu$ (see Figure~\ref{fig:drag}).

For the sake of simplicity, hereafter we assume that the control goal is to move the platoon along the circuit according to a desired plan featuring constant velocity $v_i=\nu$ and constant inter-vehicle distance $\Delta s_i=\mu$ for each unit.
In particular, the desired motion chosen for each vehicle is given by
\begin{align*}
	\xi_i =&~ \nu t-i\mu ~.
\end{align*}
Then, by defining the error with respect to the desired motion as $e_i=s_i-\xi_i$, and by denoting
\begin{align*}
	\dot{s}_i =&~ v_i = \nu+\dot{e}_i \\
	\ddot{s}_i =&~ a_i = \ddot{e}_i \\
	\Delta e_i =&~ e_{i-1}-e_i \\
	\Delta s_i =&~ \xi_{i-1}+e_{i-1}-\xi_i-e_i = \mu+\Delta e_i \\
	\Delta v_i =&~ v_{i-1}-v_i = (v_{i-1}+\nu)-(v_i+\nu) =\Delta\dot{e}_i ~,
\end{align*}
one can describe the vehicle model in terms of the displacement $e_i$:
\begin{align*}
	\ddot{e}_i =&~ \frac{1}{m}\Big(u_i-\alpha (\nu+\dot{e}_i)\big(\dot{e}_i^2-\beta \Delta e_i\big)-\gamma\Big) \\
	=&~ \frac{1}{m}\big(u_i-\alpha \nu \dot{e}_i^2-\alpha \dot{e}_i^3 \\
	&~+\nu \alpha \beta \Delta e_i+\alpha \beta \dot{e}_i\Delta e_i-\gamma\big) ~.
\end{align*}

In order to conceive a simple strategy for controlling such a platoon formation, we also assume that information or estimates of position and speed can be used to move each vehicle when alone, due to excessive distance from the others or sensor failures.
Under such an hypothesis, an intuitive approach can be developed according to the following reasoning.
\begin{itemize}
\item[--] First, the constant friction $\gamma$ can be compensated by means of a static input of the same value.
\item[--] Then, to avoid that large displacements can turn into vehicle collisions, a widely used strategy, such as the Adaptive Cruise Control (ACC, see \cite{Konrad} and references therein), can be exploited.
\end{itemize}
It is also worth of observing  that ACC in its standard formulation has the form of a Proportional-Derivative (PD) function of the inter-vehicular position, and then it can be conveniently cast in the present problem as a PD control input computed on the $\Delta e$'s.

Summing up, the chosen control input $u_i$ is designed as
\begin{align}
	\label{eq:control input}
	u_i =&~ \gamma-H_de_i-H_s\dot{e}_i \\\nonumber
	&+K_p\Delta e_i+T_p\Delta\dot{e}_i-K_f\Delta e_{i+1}-T_f\Delta\dot{e}_{i+1} ~,
\end{align}
where ACC has been set up considering both the preceding and the following vehicles.

It is worth underlining that the isolated vehicle model is
\begin{align*}
	\ddot{e}_i =&~ \frac{1}{m}\big(
	-H_de_i
	-H_s\dot{e}_i
	-\alpha \nu \dot{e}_i^2
	-\alpha \dot{e}_i^3
	\big) ~.
\end{align*}
In such a case, to assure the (local) convergence of the vehicle to the desired path, the coefficients $H_d$ and $H_s$ must be set positive.
Also, notice that this component of the controller could be able, if the position and speed information are sufficiently accurate, to solve the problem by itself.
However, completely neglecting the presence of the other vehicles is dangerous, and a collision avoidance strategy such as ACC turns out necessary.

Therefore, if the collisions are considered a primary risk or the navigation system is not perfectly reliable, it is reasonable to set up ACC in order to be at least strong as much as the rest of the control actions.
This is indeed the scenario considered in the rest of the paper.
\section{Traveling waves investigation}
Model~\eqref{eq:the model} provided with the control input~\eqref{eq:control input} may be affected by local instability because of the wake.
Indeed, the drag reduction due to this aerodynamic phenomenon is able to make the total friction less than $\gamma$.
Therefore, when the vehicle approaches the preceding one a little bit closer than $\mu$, the static component of the control input turns out bigger than the actual friction, and its effect results in increasing the forward acceleration.
Hence, if the other two components of the control input are not properly designed, the platoon may diverge from the desired formation.

In the following we develope a qualitative analysis tool, based on the PDE embedding approach described in~\cite{Innocenti}, to investigate if a chosen set of controller coefficients is compatible with the existence of traveling waves (see also~\cite{Paoletti}).

Let us introduce the embedding variable $x\in\mathbb{R}$ and the interpolating function $\xi(t,x)$, so that
\begin{align*}
	e_i(t)=&~\xi(t,x_i) \\
	\delta x=&~x_i-x_{i-1} ~.
\end{align*}
Then, $\dot{e}_{i}(t)=\partial_{t}\xi(t,x_i)$ and $\ddot{e}_{i}(t)=\partial_{tt}\xi(t,x_i)$.
Moreover, if the sought solution is sufficiently regular with respect to $x$, the following approximations can be taken into account:
\begin{align*}
	e_{i+1}(t) \approx&~ \xi(t,x_i)+\partial_{x}\xi(t,x_i)\delta x \\
	e_{i-1}(t) \approx&~ \xi(t,x_i)-\partial_{x}\xi(t,x_i)\delta x \\
	\dot{e}_{i+1}(t) \approx&~ \partial_{t}\xi(t,x_i)+\partial_{x t}\xi(t,x_i)\delta x \\
	\dot{e}_{i-1}(t) \approx&~ \partial_{t}\xi(t,x_i)-\partial_{xt}\xi(t,x_i)\delta x
\end{align*}
\begin{align*}
	\Delta e_i =&~ e_{i-1}(t)-e_i(t) \approx -\partial_{x}\xi(t,x_i)\delta x \\
	\Delta e_{i+1} =&~ e_{i}(t)-e_{i+1}(t) \approx -\partial_{x}\xi(t,x_i)\delta x \\
	\Delta \dot{e}_i =&~ \dot{e}_{i-1}(t)-\dot{e}_i(t) \approx -\partial_{xt}\xi(t,x_i)\delta x \\
	\Delta \dot{e}_{i+1} =&~ \dot{e}_{i}(t)-\dot{e}_{i+1}(t) \approx -\partial_{xt}\xi(t,x_i)\delta x ~.
\end{align*}
Substituting the above quantities into the single vehicle equation and removing the index $i$, since all the platoon units are the same, one obtains the PDE model
\begin{align*}
	\partial_{tt}\xi =&~ \frac{1}{m}\big(
	-H_d\xi
	-H_s\partial_{t}\xi
	-\delta x(K_p-K_f)\partial_{x}\xi \\&
	-\delta x(T_p-T_f)\partial_{xt}\xi
	-\alpha \nu \partial_{t}\xi^2
	-\alpha \partial_{t}\xi^3 \\&
	-\nu \alpha \beta \delta x\partial_{x}\xi
	-\alpha \beta \delta x\partial_{t}\xi\partial_{x}\xi
	\big) ~.
\end{align*}
In order to investigate the existence of traveling waves, the moving coordinate
\begin{align*}
	\zeta=&~ct+kx
\end{align*}
is introduced, where $c$ and $k$ are referred to as \emph{angular frequency} and \emph{wave number}:
\begin{align*}
	\xi(t,x) =&~ \xi(\zeta) \\
	\partial_{t}\xi(t,x) =&~ c\partial_{\zeta}\xi(\zeta) = c\dot{\xi}(\zeta) \\
	\partial_{x}\xi(t,x) =&~ k\partial_{\zeta}\xi(\zeta) = k\dot{\xi}(\zeta) \\
	\partial_{tt}\xi(t,x) =&~ c^2\ddot{\xi}(\zeta) \\
	\partial_{xt}\xi(t,x) =&~ ck\ddot{\xi}(\zeta) ~.
\end{align*}
By substituting the above quantities into the PDE model, one finds the so called \emph{reference ODE} (see~\cite{Innocenti})
\begin{align}
\label{eq:reference ODE}
	 \ddot{\xi}
	+a\dot{\xi}
	+b\xi =&~
	-p\dot{\xi}^2
	-q\dot{\xi}^3 ~,
\end{align}
where the following coefficients have been introduced for the sake of simplicity
\begin{align*}
	K =&~ \frac{K_p-K_f}{\nu\alpha \beta } \\
	T =&~ \frac{T_p-T_f}{m} \\
	\varrho =&~ \delta xk  \\
	a =&~ \frac{cH_s+\varrho \nu\alpha \beta (K+1)}{mc(c+\varrho T)} \\
	b =&~ \frac{H_d}{mc(c+\varrho T)} \\
	p =&~ \frac{\alpha (\nu c+\beta \varrho )}{m(c+\varrho T)} \\
	q =&~ \frac{\alpha c^2}{m(c+\varrho T)} ~.
\end{align*}

System~\eqref{eq:reference ODE} must now be investigated in search of a periodic solution $\xi(\zeta+\tau)=\xi(\zeta)$.
To this aim, we exploit a standard bifurcation analysis approach.
In particular, since~\eqref{eq:reference ODE} has order two, we can look for possible limit cycles encircling the equilibrium in $\xi=\dot{\xi}=0$ when this latter turns unstable.
Then, we just enforce a Hopf bifurcation scenario by choosing
\begin{align}
\label{eq:Hopf conditions}
	a<0 ~, \quad a^2<4b ~.
\end{align}
Hence, for each (small) $\varepsilon>0$, we obtain the possible \emph{dispersion curve}
\begin{align}
\label{eq:dispersion curve}
	\varrho(c)  =&~ \frac{cH_s-\varepsilon mc^2}{\varepsilon mcT-\nu \alpha \beta (K+1)} ~,
\end{align}
valid if $c+\varrho(c) T\neq 0$.
It is worth stressing that, when the dispersion curve of the PDE model is brought back to the original platoon, it boils down to a set of points $(c,\varrho(c))$, since only certain wave numbers are compatible with the number $N$ of vehicles in the platoon, i.e. the platoon length $N\delta x$ with respect to the embedding variable $x$ (see~\cite{Innocenti,Paoletti} for further details). 

Moreover, observe that conditions~\eqref{eq:Hopf conditions} do not guarantee the existence of a limit cycle by themselves, since the Hopf bifurcation can happen in two variants, namely the super- and the sub-critical cases (see, e.g.,~\cite{Marsden}).
Therefore, the actual existence of the limit cycle in the reference ODE model must be checked with other tools, such as numerical simulations.
\section{Numerical example}
\begin{figure}[tb]
\begin{center}
\includegraphics[width=0.70\columnwidth]{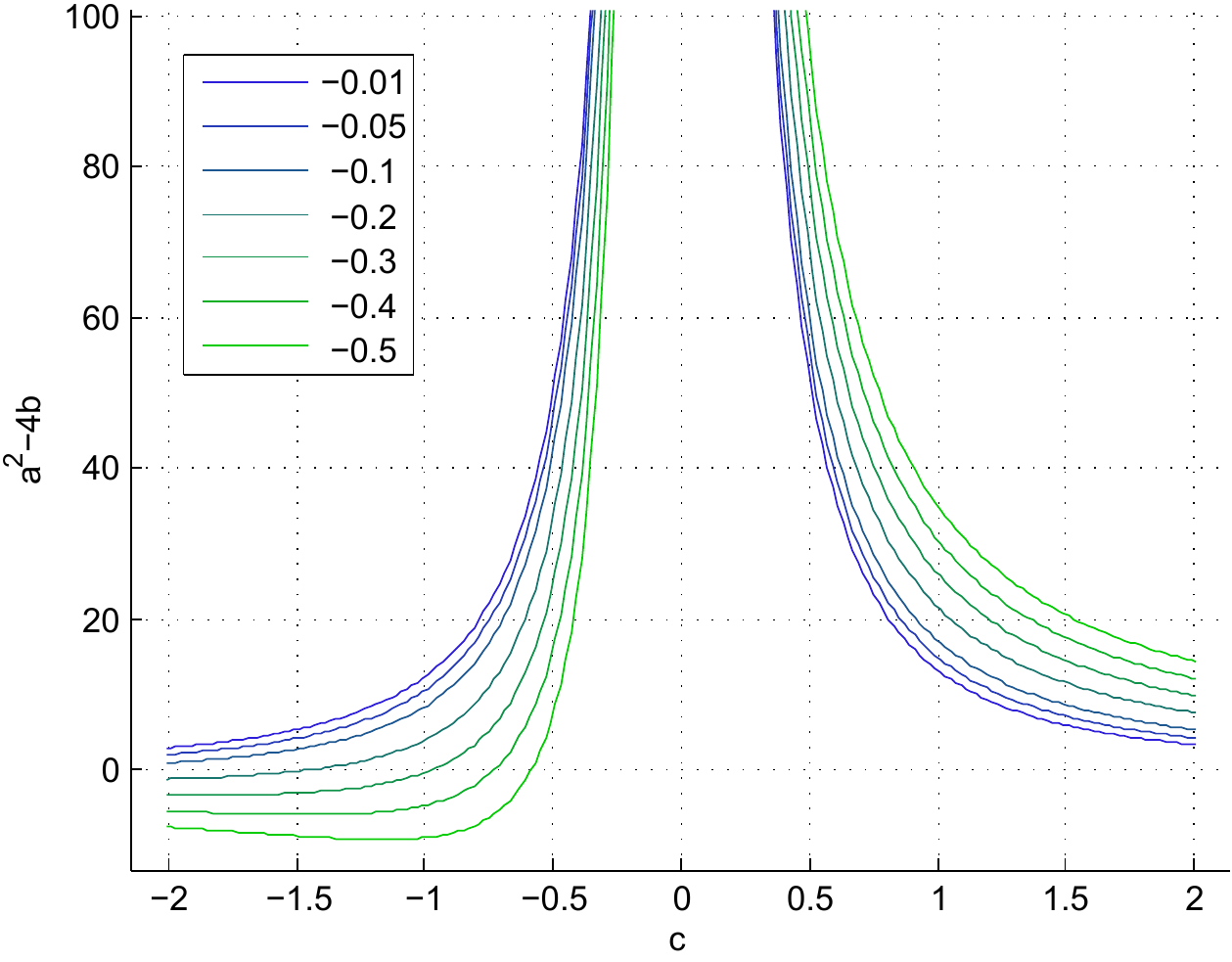}
\end{center}
\caption{Values assumed by the quantity $a^2-4b$ in the first scenario for different values of $\varepsilon$.
\label{fig:delta quadro}} 
\end{figure}
\begin{figure}[tb]
\begin{center}
\includegraphics[width=0.70\columnwidth]{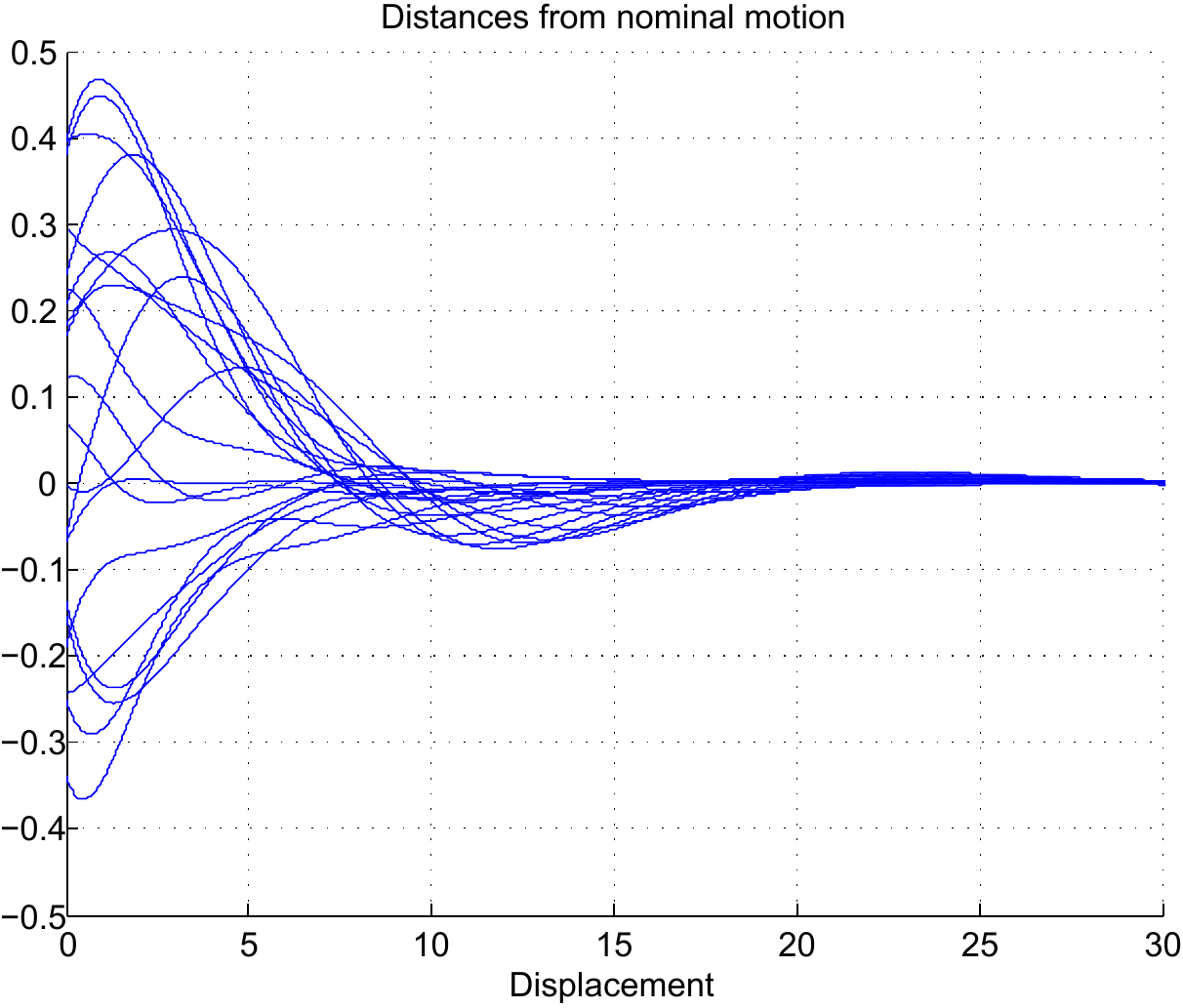}
\end{center}
\caption{Vehicle displacements in the first scenario.
\label{fig:e 4}} 
\end{figure}
\begin{figure}[tb]
\begin{center}
\includegraphics[width=0.70\columnwidth]{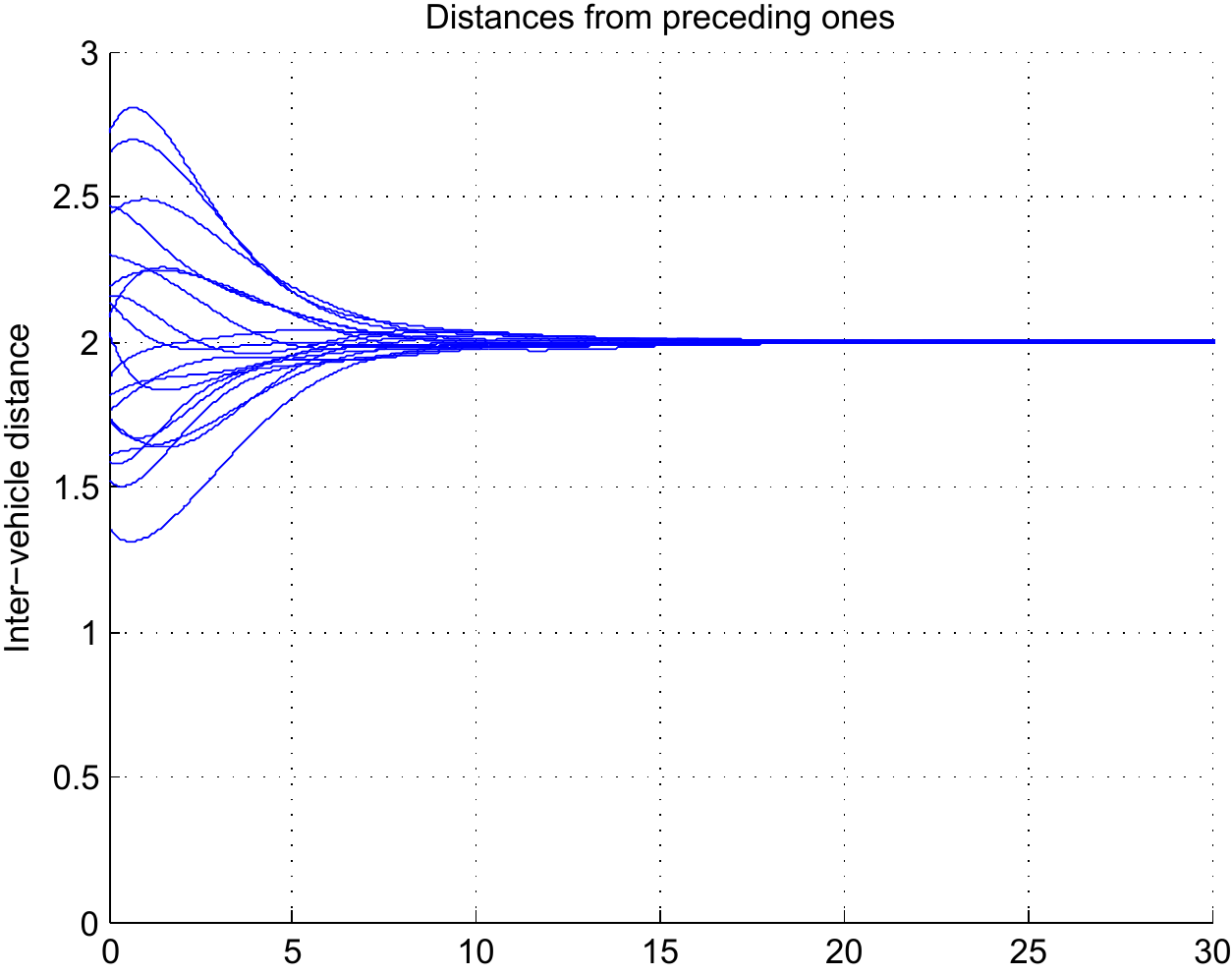}
\end{center}
\caption{Inter-vehicle distances in the first scenario.
\label{fig:delta e 4}} 
\end{figure}
\begin{figure}[tb]
\begin{center}
\includegraphics[width=0.70\columnwidth]{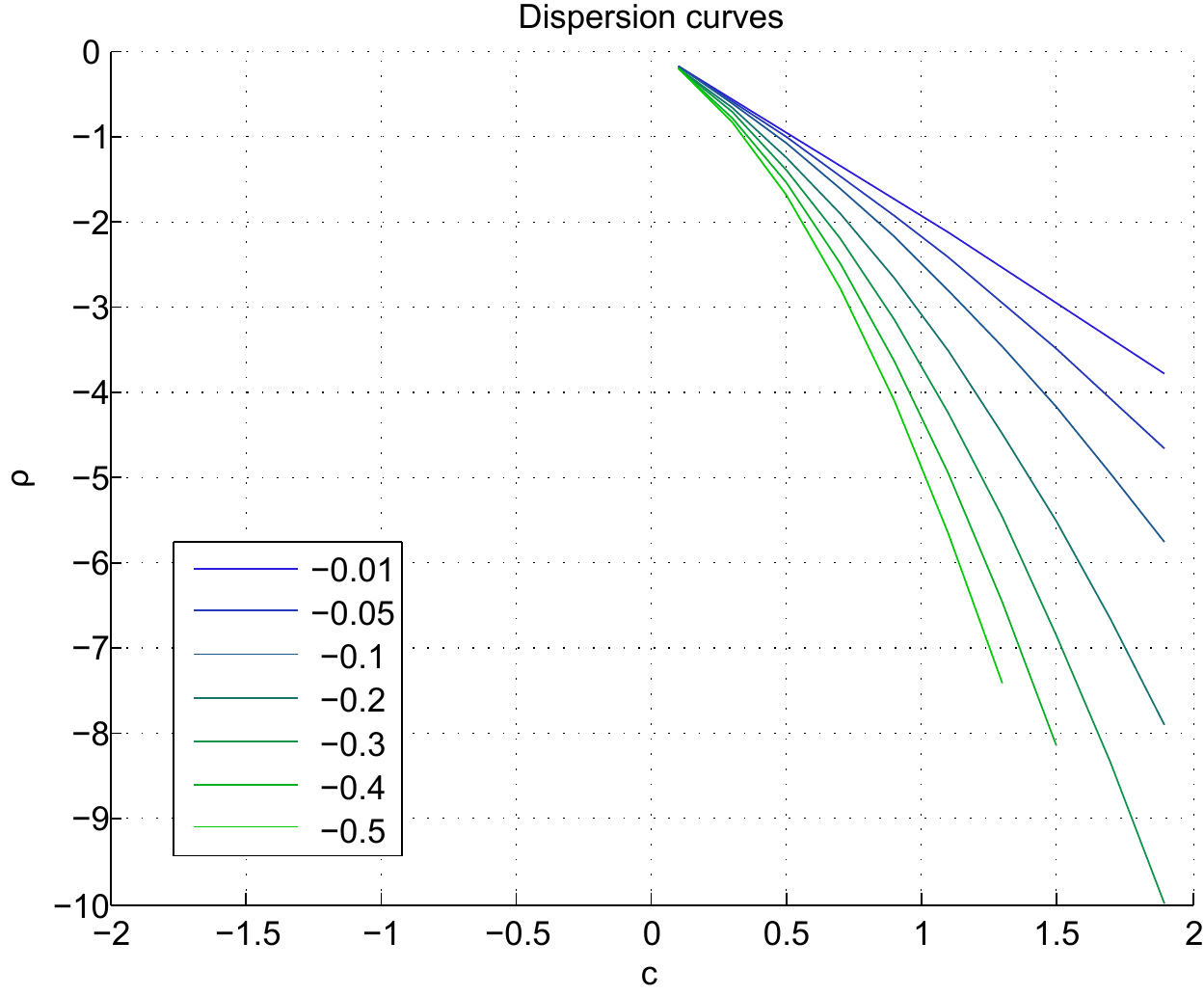}
\end{center}
\caption{Dispersion curves of the second scenario for different values of $\varepsilon$. Parameter $c$ varies in the range $(-2,2)$.
\label{fig:dispersion curves}} 
\end{figure}
\begin{figure}[tb]
\begin{center}
\includegraphics[width=0.70\columnwidth]{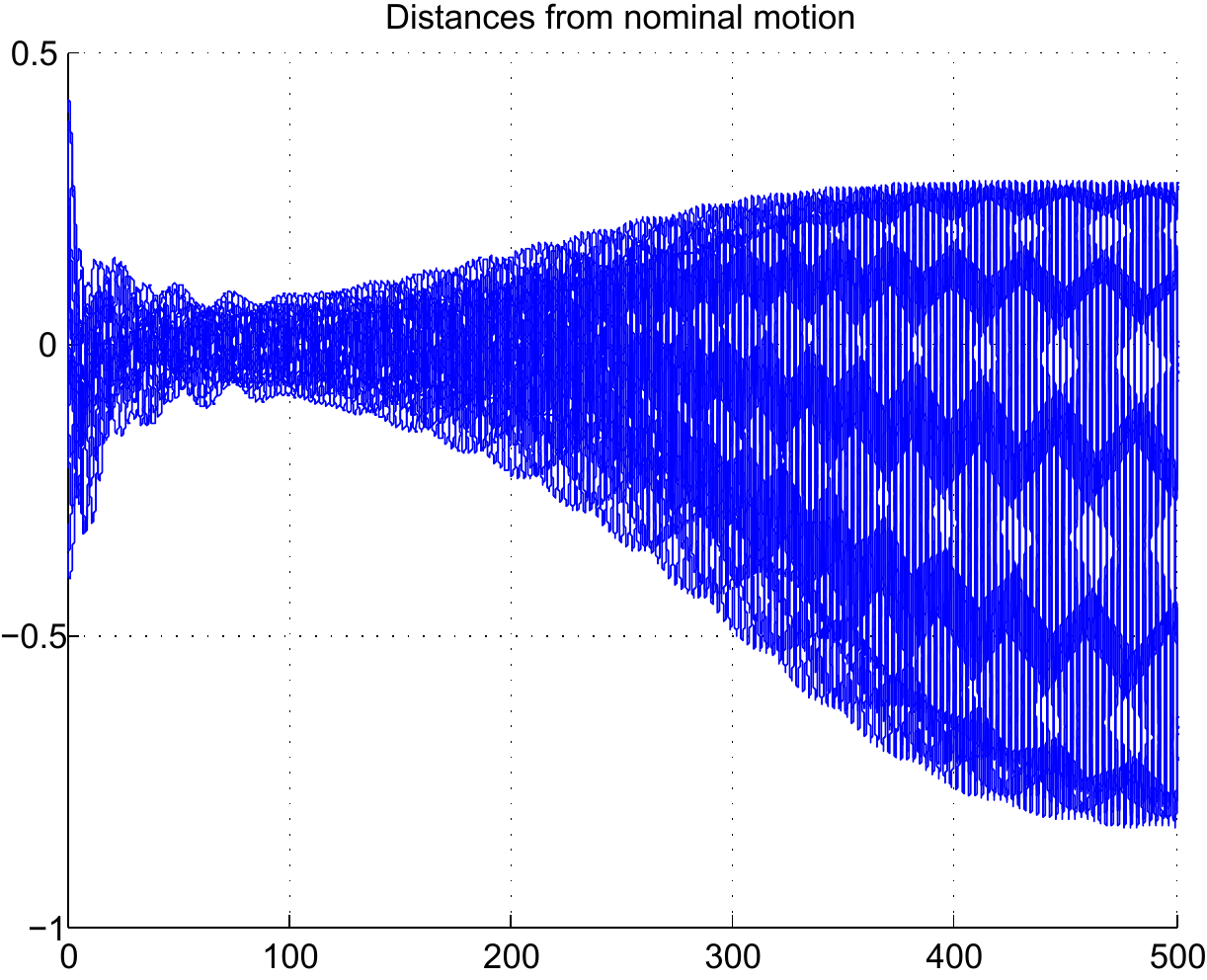}
\end{center}
\caption{The traveling wave found in the second scenario.
\label{fig:e 3}} 
\end{figure}
\begin{figure}[tb]
\begin{center}
\includegraphics[width=0.70\columnwidth]{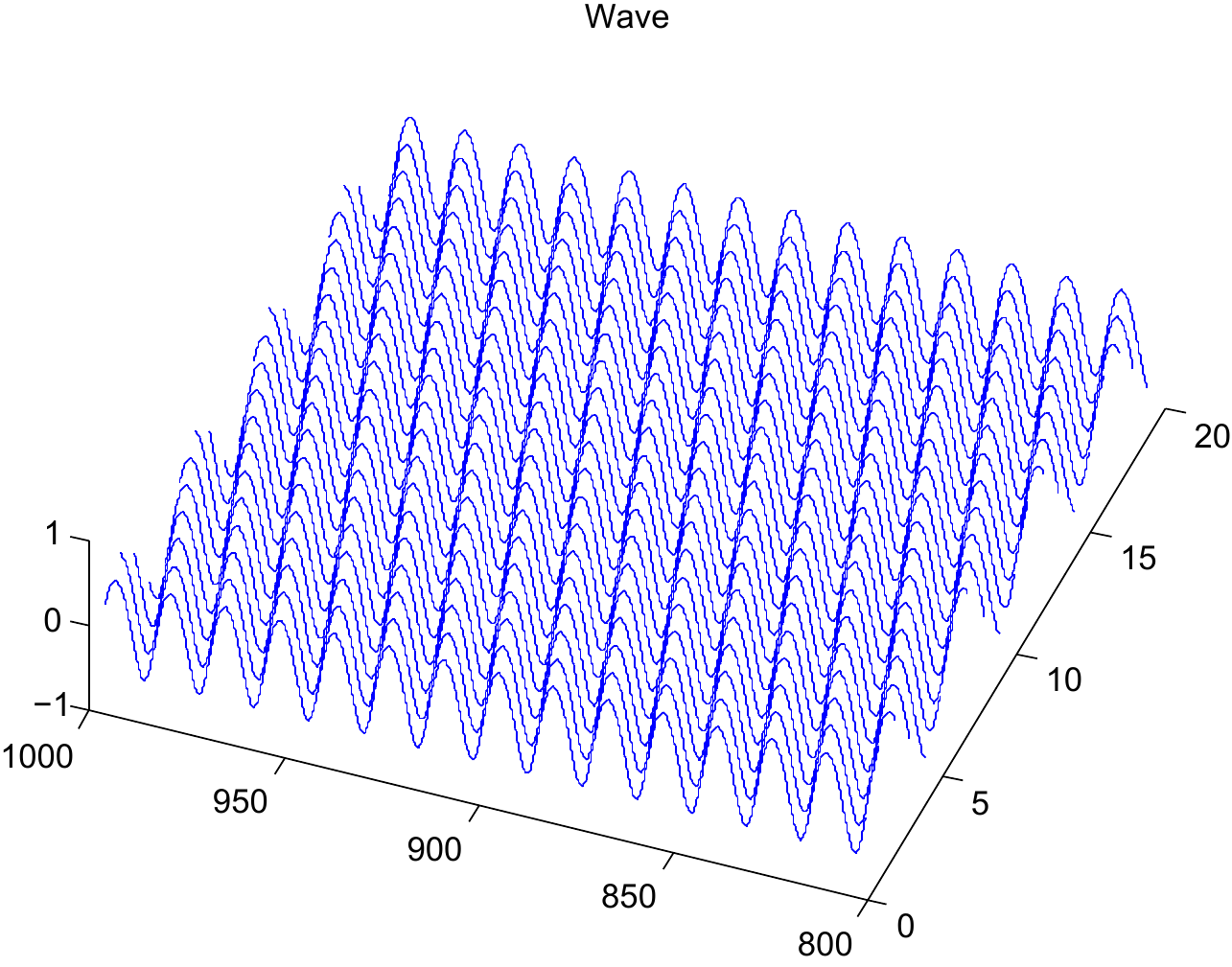}
\end{center}
\caption{The traveling wave found in the second scenario.
\label{fig:3D wave}} 
\end{figure}
\begin{figure}[tb]
\begin{center}
\includegraphics[width=0.70\columnwidth]{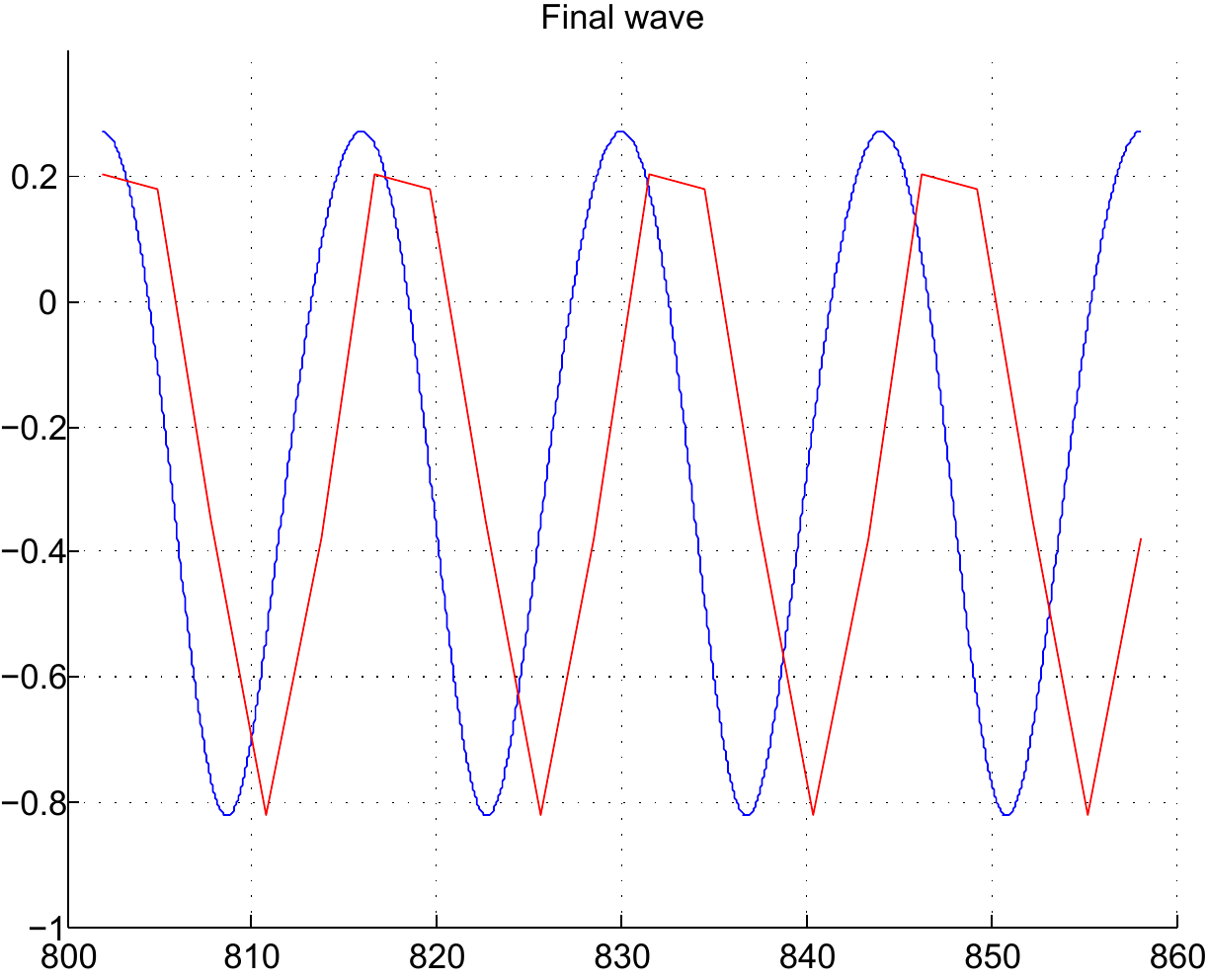}
\end{center}
\caption{Comparison between the displacement from the desired motion of a single vehicle (blue) after the transient, and the spatial profile of the platoon displacements at the end of the simulation (red). The two graphs have been aligned for the sake of clarity.
\label{fig:wave profile}} 
\end{figure}
\begin{figure}[tb]
\begin{center}
\includegraphics[width=0.70\columnwidth]{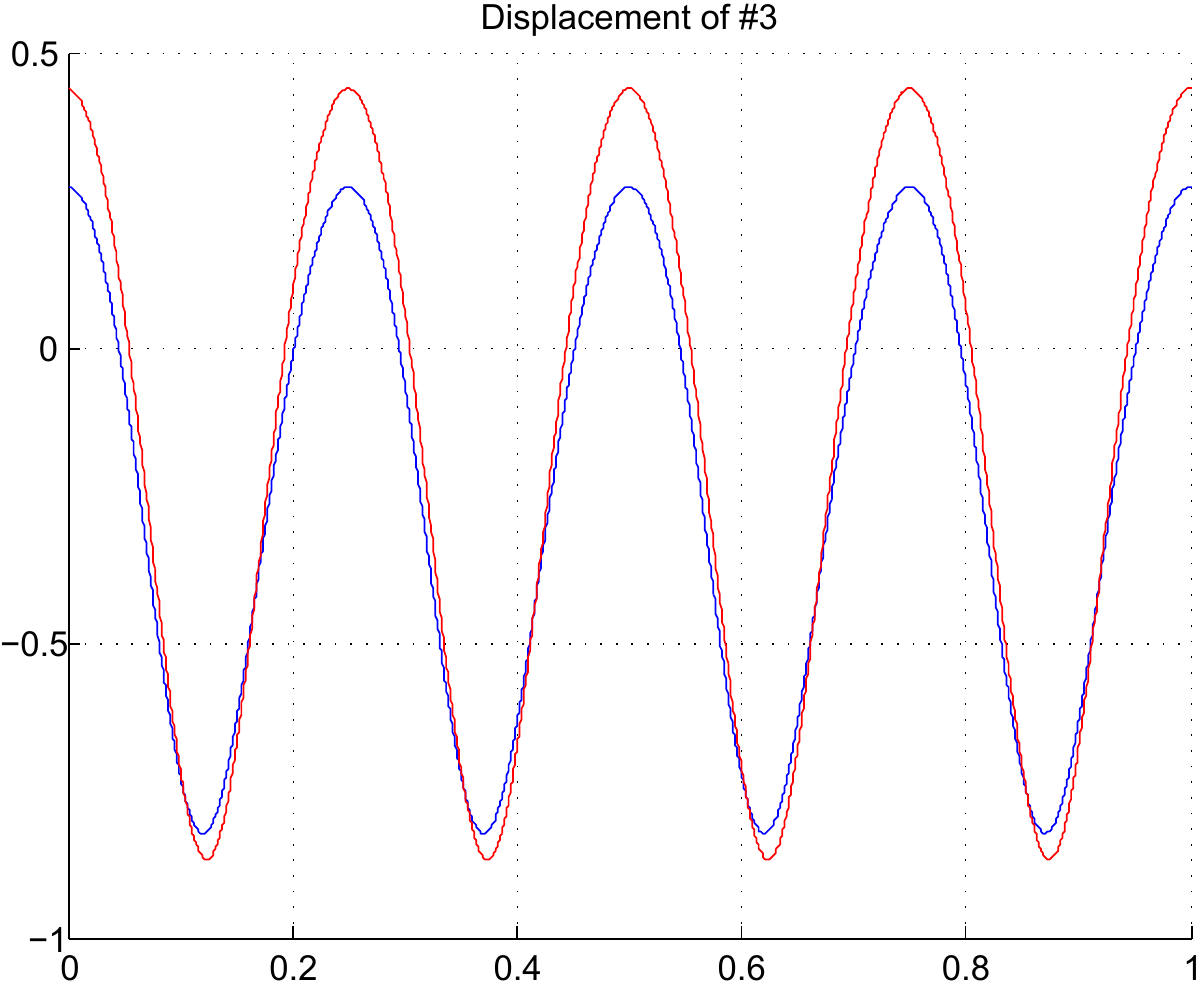}
\end{center}
\caption{Comparison between the displacement from the desired motion of a single vehicle (blue) after the transient, and the spatial solution of the reference ODE (red). The two graphs have been aligned for the sake of clarity.
\label{fig:ODE profile}} 
\end{figure}
In this section a toy model (not related to any real world platoon) is used for the sole purpose of illustrating the tools developed in the previous section to investigate the existence of traveling waves in a jammed circular platoon.
The vehicle parameters are
\begin{align*}
	m = 1 ~,\quad
	\alpha = 1.0 ~,\quad
	\beta = 0.1 ~,\quad
	\gamma = 1 ~,
\end{align*}
while the objective formation is characterized by
\begin{align*}
	\mu = 2 ~, \quad \nu = 1 ~.
\end{align*}
The coefficients of the control input component based on the navigation system are
\begin{align*}
	H_d = 0.10 ~,\quad
	H_s = 0.30 ~,
\end{align*}
privileging the information/estimate of the speed over the position.
Notice, this is a common situation in real world vehicles.
In such a framework, we want to investigate if the addition of a ACC strategy may induce traveling waves in the platoon.

As first scenario let us consider the following parameters for the ACC controller: 
\begin{align*}
	K_p = 0.06 ~,\quad
	T_p = 0.55 ~,\quad
	K_f = 0 ~,\quad
	T_f = 0 ~.
\end{align*}
Even if the Hopf conditions~\eqref{eq:Hopf conditions} are satisfied for certain values of $c$ and the related $\varrho$, see Figure~\ref{fig:delta quadro}, numerical simulations of the reference ODE system~\eqref{eq:reference ODE} exclude the existence of limit cycles.
Therefore, according to our previous analysis, we do not expect the platoon to exhibit traveling waves when the ACC control input is configured with the above parameters.
Figure~\ref{fig:e 4} and Figure~\ref{fig:delta e 4} show one of many similar numerical simulations obtained for random starting conditions close to the desired motion of the platoon:
Each vehicle reaches the desired position, and all the inter-vehicle distances $\Delta e_i$ remain positive.

As second scenario, let us configure ACC with the following parameters:
\begin{align*}
	K_p = 0.06 ~,\quad
	T_p = 0.01 ~,\quad
	K_f = 0 ~,\quad
	T_f = 0 ~.
\end{align*}
Introducing the above numbers in~\eqref{eq:Hopf conditions} we obtain a number of possible Hopf bifurcation scenarios, where, this time, the reference ODE shows actual limit cycles in numerical simulations.
The dispersion curves in Figure~\ref{fig:dispersion curves} are derived from~\eqref{eq:dispersion curve} for $\varepsilon$ ranging from $0.01$ to $0.50$.
Their graph is restricted to the cases in which a limit cycle exists.

Therefore, we expect the platoon to show traveling waves in this second scenario.
In particular, since $\varrho$ turns always out negative, the wave is supposed to move backward along the platoon, just as one would expect from the asymmetric configuration of ACC, that allows a perturbation to move from a vehicle to the following one, but not to the preceding.
Generally speaking, we also expect the wave to depend on the vehicles number.
It is also worth stressing that the platoon could be able to sustain multiple waves, since nonlinear systems are not limited to a single stationary solution.
However, the developed tool does not provide any information on the stability of each possible wave, that in turn could not be attractive for the neighbor trajectories.
To check for waves existence we again rely on numerical simulation.
Figure~\ref{fig:e 3} illustrates a platoon of $20$ vehicles initialized in random conditions close to the desired motion.

In less than $500$ time steps the system trajectory converges to the traveling wave highlighted in Figure~\ref{fig:3D wave}.

The spatial profile of the wave at the end of the simulation is reported in Figure~\ref{fig:wave profile}, and it shows that the spatial period comprises $4$ time periods, that is $nc\phi=N\varrho$, $\phi$ being the time period and $n=4$.
Observe that $\varrho/c=n\phi/N=-2.81$ is compatible with the computed dispersion curves, but it states that a time interval equal to $2.81$ along the temporal wave corresponds to a single unit (vehicle) in the spatial wave.
Hence, it suggests that for a time period equal to $14.05$, as in this case, the solution of the reference ODE may be just a raw approximation of the actual platoon wave, because of the little number of units per period.

In Figure~\ref{fig:ODE profile} the periodic motion of a single vehicle is compared with the periodic solution of the reference ODE for $\varepsilon=-0.03$ and $c=0.95$.

\section{Conclusions}
In this paper we have introduced a novel nonlinear model for describing a platoon of identical terrain vehicles, moving in a circuit.
Each unit has been assumed subjected to a constant friction and to a nonlinear drag featuring aerodynamic effects depending also on the wake from the preceding one.
The desired formation consisted of evenly distributed vehicles moving at constant speed.
Each unit has also been assumed to have a minimal knowledge about its own positioning along the circuit, as well as about the inter-vehicle distances with the preceding and following units.
Moreover, a ACC controller has been set up according to two different configurations, and the possible rising of complex phenomena has been investigated by mean of bifurcation analysis tools exploiting a recent embedding technique independent from the vehicles number.
The results show that a wrong configuration of the ACC controller can drive the platoon formation to instability and to the rise of self-sustained traveling waves.
\balance
\bibliographystyle{apalike}

%

%
%
\end{document}